\documentclass{elsart4-1}

\usepackage{epsfig}
\usepackage{float}
\usepackage[compress]{cite}
\usepackage[latin1]{inputenc}
\usepackage{amssymb}

\usepackage[english,francais]{babel}

\newtheorem{e-proposition}[theorem]{Proposition}

\newtheorem{e-definition}[theorem]{Definition\rm}

\def\og{\leavevmode\raise.3ex\hbox{$\scriptscriptstyle\langle\!\langle$~}}
\def\fg{\leavevmode\raise.3ex\hbox{~$\!\scriptscriptstyle\,\rangle\!\rangle$}}

\begin{document}

\centerline{Astrophysics}
\begin{frontmatter}


\selectlanguage{english}
\title{The flux suppression at the highest energies}


\selectlanguage{english}
\author{Diego Harari}
\ead{harari@cab.cnea.gov.ar}

\address{Centro At\'omico Bariloche and Instituto Balseiro\\
San Carlos de Bariloche, Argentina}



\begin{abstract}

Almost half a century ago, Greisen, Zatsepin and Kuz'min (GZK) predicted a ``cosmologically meaningful termination" of the spectrum of cosmic rays at energies around $10^{20}$ eV due to their interaction with the cosmic microwave background, as they propagate from distant extragalactic sources. A suppression of the flux above $4\times 10^{19}$ eV is now confirmed. We argue that current data are insufficient to conclude whether the observed feature is due to energy-loss during propagation, or else because the astrophysical accelerators reach their limit, or indeed to a combination of both source properties and propagation effects. We discuss the dependence of the spectral steepening upon the cosmic-ray composition, source properties, and intervening magnetic fields, and speculate on the additional information that may be necessary to reach unambiguous conclusions about the origin of the flux suppression and of the mechanisms behind the acceleration of cosmic rays up to the highest observed energies.     

{\it To cite this article: D. Harari, C. R. Physique  (2014).}

\vskip 0.5\baselineskip

\selectlanguage{francais}
\noindent{\bf R\'esum\'e}
\vskip 0.5\baselineskip
\noindent
{\bf La suppression du flux aux plus hautes \'energies}

Voilà presque un demi-siècle, Greisen, Zatsepin et Kuz'min (GZK) prédisaient, à  des énergies d'environ $10^{20}$~eV, une ``fin d'origine cosmologique" du spectre des rayons cosmiques en raison de leurs interactions avec le fond diffus cosmologique le long de leur parcours depuis des sources extragalactiques lointaines. La suppression du flux au-delà de  $ 4 \times 10^{19}$ eV est aujourd'hui confirmée, cependant les données actuelles sont insuffisantes pour conclure sur son origine :  Est-elle due  à une perte d'énergie lors de la propagation ? À la puissance limitée des accélérateurs astrophysiques ? Ou bien à une  combinaison des propriétés des  sources et des effets de propagation ?  Nous présentons les relations entre l'infléchissement spectral observé et la composition des rayons cosmiques,  les propriétés de leurs sources, et la présence de champs magnétiques. Nous discutons des informations complémentaires  nécessaires pour conclure sans ambiguïté sur l'origine de la suppression du flux et sur les mécanismes à l'origine de l'accélération des rayons cosmiques aux énergies les plus élevées observées.

{\it Pour citer cet article~: D. Harari, C. R.
Physique  (2014).}

\keyword{Cosmic rays; Ultra-high energies; GZK } \vskip 0.5\baselineskip
\noindent{\small{\it Mots-cl\'es~:} Rayons cosmiques; Ultra-hautes \'energies; GZK}}
\end{abstract}
\end{frontmatter}

\selectlanguage{francais}

\selectlanguage{english}
\section{Introduction}
\label{}

The existence of about 400 photons with median energy $6\times 10^{-4}$ eV in each cubic centimeter everywhere throughout the Universe    
has a strong impact upon the propagation of cosmic rays with energy around 23 orders of magnitude larger. This was first noted by
Greisen \cite{Greisen1966} and independently by Zatsepin and Kuz'min \cite{Zatsepin1966}, soon after the discovery of the cosmic microwave background (CMB) \cite{Penzias65}.
They realized that protons with energy around and above $6\times 10^{19}$ eV can produce pions in their collisions with the CMB photons, and 
estimated the timescale for energy loss to be several hundred times smaller than the expansion time of the Universe. This led them to predict a 
``cosmologically meaningful termination" of the spectrum of cosmic rays (CRs). If the sources are uniformly distributed across the Universe, the ``GZK effect" implies a strong flux suppression. Only a relatively small local neighborhood (in cosmological terms) contributes significantly to the flux of protons that arrive at Earth with energy above the threshold for pion-photoproduction, while the entire Hubble volume fills up the flux at lower energies.  
Even the one event recorded with the Volcano Ranch array \cite{Linsley1963} at $10^{20}$ eV three years before the GZK prediction appeared surprising.
A strong flux suppression was also predicted if the cosmic rays were heavier nuclei instead of protons, since photodisintegration also occurs 
above a comparable energy threshold over cosmologically short timescales. 
   
Forty-two years after the GZK prediction, the High Resolution Fly's Eye experiment (HiRes) \cite{HiRes2008} and the Pierre Auger Observatory \cite{Auger2008} independently measured a suppression of the flux of cosmic rays above $4\times 10^{19}$ eV compatible with a ``cosmologically meaningful termination". The efforts to improve accuracy and exposure to a flux at the level of 1 particle per ${\rm km}^2$ per century were largely driven by results from the AGASA Observatory, that gave unconfirmed evidence for trans-GZK particles \cite{Takeda2003}.  Updated results from Auger \cite{Letessier2013} indicate that at $4\times 10^{19}$ eV the flux is half of what would be expected with a power-law extrapolation from smaller energies, and the significance with which the suppression is confirmed is more than $20\sigma$. The Telescope Array (TA) \cite{TAspectrum,Kido2013} has now added independent confirmation of the suppression. The features observed by the different experiments are compatible within their systematic uncertainties \cite{SpectrumWG2012}. 

In this article we will first summarize some aspects of the GZK effect. Then we will argue that, striking as it may be that the observed suppression matches the features predicted almost half a century ago, the physical origin of the suppression remains uncertain. The spectral steepening depends upon a variety of yet poorly constrained parameters, such as the CR composition, the nature and distribution of the sources, the range of maximum energy to which they can accelerate CRs, the sources spectral index and evolution, and the amount of diffusion across intervening magnetic fields. Current data are insufficient to conclude whether the flux is suppressed due to an energy-loss propagation effect, or else because astrophysical accelerators reach their maximum power at energies comparable to the GZK threshold, or indeed to a combination of both source physical properties and propagation effects. Degeneracy in the parameter space may prevent the resolution of this puzzle with measurements of the spectral shape alone.

\section{The GZK horizon for protons and nuclei}

In the rest frame of a proton with energy around $10^{20}$ eV, most incoming CMB photons have energies above 150 MeV, sufficient to produce the $\Delta^+$ resonance, that decays to a neutron and a $\pi^+$ or to a proton and a $\pi^0$: 
\begin{equation}
p+\gamma_{\rm CMB}\rightarrow \Delta^+ \rightarrow p + \pi^0 \quad  {\rm or}\quad  n + \pi^+ \ \ .
\end{equation}
An order of magnitude estimate of the energy attenuation length can be obtained by approximating the interaction cross-section to be $\sigma\approx 10^{-28} {\rm cm}^2$ and that there are about 400 CMB photons per cubic centimeter,  which leads to a mean free path of the order of 10 Mpc. Since a proton loses on average 20\% of its energy in each interaction, its energy will decrease by an order of magnitude after traversing 100 Mpc. Taking into account the spectral distribution of CMB photons, the effective threshold for the GZK effect is actually lower, of the order of $3\times 10^{19}$ eV. The energy of the protons is also attenuated by $e^+e^-$ pair production, a process that becomes dominant at lower energies. The energy is also adiabatically red-shifted, a process that becomes relevant over cosmological travel times. Detailed calculations must also account for the spectral distribution and redshift of the CMB photons. The interaction with the infrared extragalactic background light can be neglected in the case of protons, but not for heavy nuclei, that lose energy in photodisintegration processes. When the center of mass energy exceeds the giant dipole resonance, nucleons can be emitted. Less energy is required in the center of mass, as compared to pion-photoproduction, and both the CMB and infrared photons are relevant targets. 

After the first initial estimates and predictions by Greisen and by Zatsepin and Kuz'min, several detailed calculations of energy loss due to propagation effects across the universal photon fields were performed, both for protons (e.g., \cite{Stecker1968,Hill1985,Berezinsky1988,Berezinsky1990,Aharonian1990,Yoshida1993,Rachen1993,Aharonian1994,Waxman1995,Stanev2000,
Berezinsky2005,Berezinsky2006,
Kachelriess2009}) as well as for heavier nuclei (e.g., \cite{Puget1976,Epele1998,Stecker1999,Khan2005,Allard2005,Allard2006,Harari2006,Hooper2007,Allard2008,Hooper2008,Allard2009}). Some recent reviews are \cite{Watson2013,Kotera2011,Stanev2009}. Public codes are available \cite{Kampert2013,Aloisio2012} to realistically simulate CR propagation taking into account all relevant particle interactions, tools that become more necessary to discriminate between alternative scenarios as the measurements improve.

\begin{figure}[h]
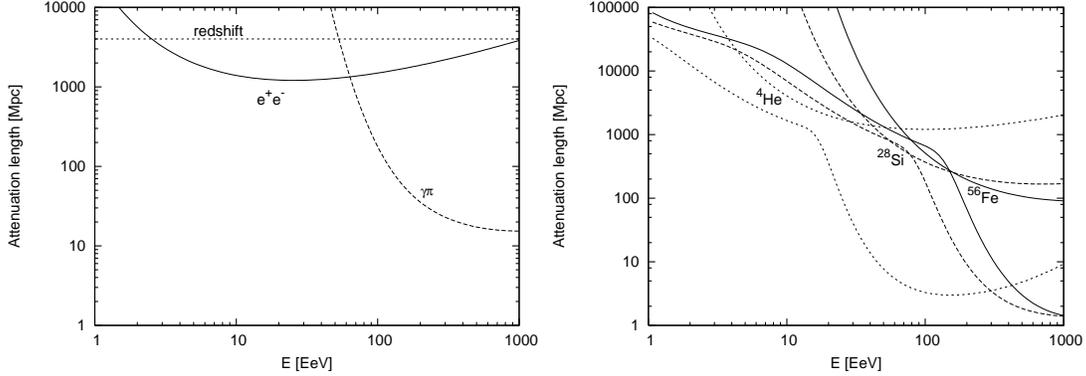

\centerline{\epsfig{file=attlengthp.ps,width=2in,angle=-90}\epsfig{file=attlength.ps,width=2in,angle=-90}}
\vspace*{8pt}
\caption{Attenuation lengths as a function of energy for protons (left) and for different nuclei (right, with dotted lines used for $^4$He, dashed lines for $^{28}$Si and solid lines for $^{56}$Fe) (taken from \cite{Harari2006}). In the case of nuclei lower curves
are due to photodisintegration and upper curves to pair production.  Energy is expressed in units of EeV=$10^{18}$ eV.
\label{fig:attenuation}}
\end{figure}

Here we will illustrate some generic features of the GZK effect following the calculations in \cite{Harari2006}. Figure \ref{fig:attenuation} displays the results for the energy attenuation length for protons (left) and for nuclei (right). 
Energy loss processes limit the distance from which a source can contribute significantly to the flux at Earth. The size of the ``GZK horizon" is estimated as follows. Under a continuous energy loss approximation, particles are assumed to be injected with an input spectrum  $\frac{dN}{dE}\propto  E^{-\alpha}$, and an attenuation factor is evaluated, given by the fraction of the events injected with energy above $E_{th}$ which still remain with $E > E_{th}$ 
after traversing a distance $D$.  Assuming that the sources are uniformly distributed and have similar intrinsic CR luminosities and spectra, the fraction of the events observed above a given energy threshold which originated in sources farther away than a distance $D$ can be derived by integration of the attenuation factor. 
\begin{figure}[h]
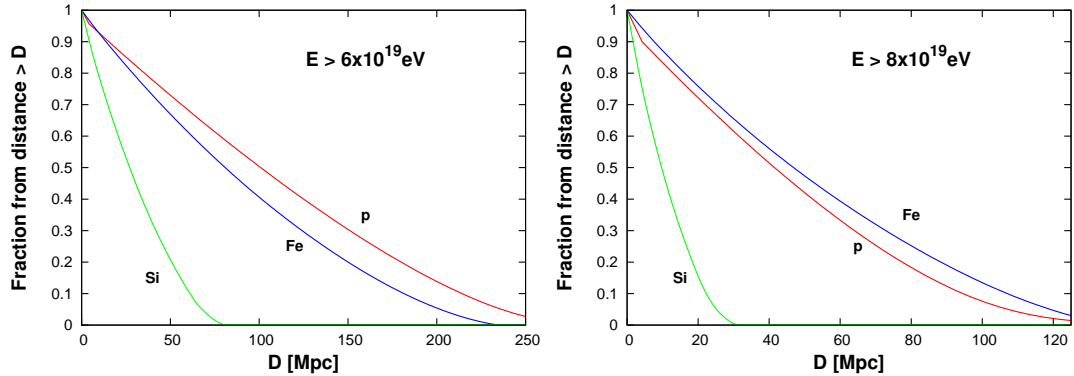

\centerline{\epsfig{file=gzk60.ps,width=2.in,angle=-90}\epsfig{file=gzk80.ps,width=2.in,angle=-90}}
\vspace*{8pt}
\caption{GZK horizon: fraction of CRs that arrive at Earth with energy above $6\times 10^{19}$ eV (left panel) and  $8\times 10^{19}$ eV (right panel),
from sources that are farther away than a distance $D$ and that inject protons, silicon and iron, respectively (assuming a uniform distribution of sources with equal intrinsic luminosity and continuous energy loss). Around 90\% of the flux of protons with $E > 6\times 10^{19}$~eV should come from distances smaller than 200 Mpc. This distance shrinks to 90 Mpc if $E > 8\times 10^{19}$~eV. The ``horizon" is of similar size for iron nuclei, and is smaller for intermediate-mass nuclei. Results have some dependence upon the injection spectral index, and are shown here for $\alpha = 2.2$.
\label{fig:horizons}}
\end{figure}
The result is illustrated in figure \ref{fig:horizons}, for energy thresholds of $6\times 10^{19}$ eV (left panel) and 
$8\times 10^{19}$ eV (right panel), for sources that inject protons, silicon and iron, respectively. For instance, approximately 90\% of the flux of protons with $E > 6 \times 10^{19}$ eV should come from distances smaller than 200 Mpc if the sources were distributed uniformly across the Universe.
The ``GZK horizon" is of comparable size for protons and iron nuclei, and is smaller for intermediate-mass nuclei. It decreases for higher energy thresholds. 

Note that even if a single nuclear mass type were injected at the sources, the distribution of the masses of the CRs arriving at Earth would be quite wide, with contributions from many different nuclear species. An important qualitative difference between the case of injected proton or nuclei is that the energy-loss mechanism for protons leads to a ``pile-up"\cite{Hill1985,Berezinsky1988,Berezinsky2005,Berezinsky2006} of particles below the threshold for pion-photoproduction. Instead the fragments from nuclei photodisintegration are typically also above threshold, and hence no significant pile-up occurs at these energies. Another relevant difference is the amount of ``cosmogenic" neutrinos that are produced as a side-effect of CR propagation \cite{Berezinsky1969,Stecker1979,Hill1985,Engel2001}, which is larger for protons than for heavy nuclei \cite{Hooper2005}. 

\section{Measurements of the flux suppression}

\begin{figure}[h]
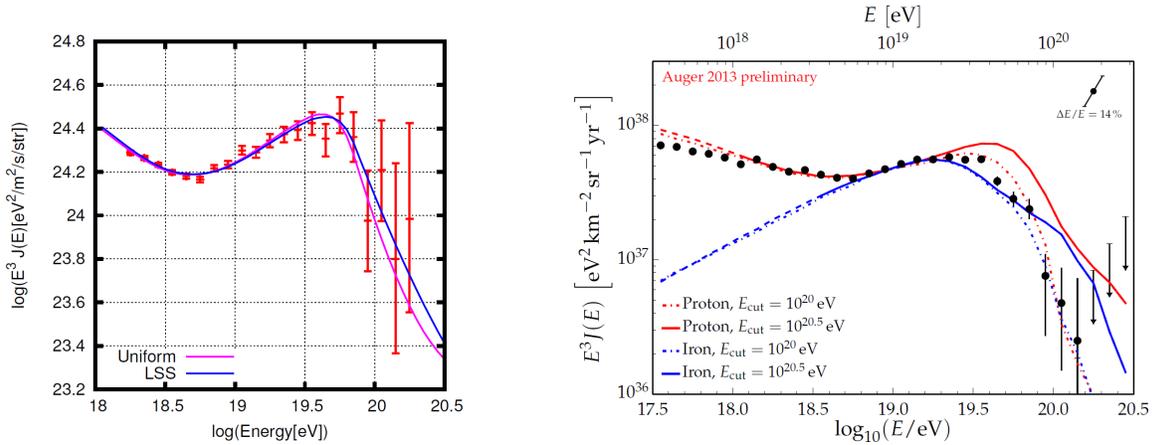

\centerline{\epsfig{file=TAfit.eps,width=2.5in}\hspace{0.5in}\epsfig{file=Augerfit.eps,width=3.in}}
\vspace*{8pt}
\caption{Recent measurements of the flux (rescaled by $E^3$) of CRs at the highest energies and comparison to the flux-suppression features expected in some simple astrophysical scenarios. Left: TA measurements and fits to a model with extragalactic proton sources, assuming a uniform distribution (pink line) or a distribution following the large-scale structure (blue line). The spectral index $\alpha$ and the source evolution factor (luminosity is assumed to scale with redshift $z$ as $(1+z)^m$) are best-fitted by $\alpha=2.36, m=4.5$ and $\alpha=2.39, m=4.4$ respectively (taken from \cite{Kido2013}). Right: Auger measurements and comparison to models based on pure proton, with $\alpha=2.35$ and $m=5$ (red) or pure iron, with $\alpha=2.3$ and $m=0$ (blue) extragalactic sources, with an injection exponential cut-off around $10^{20.5}$ eV (solid lines) and $10^{20}$ eV (dotted lines) (taken from \cite{Letessier2013}).
\label{fig:spectra}}
\end{figure}

If the sources of CRs with the highest energies lie in astrophysical sites spread over the Universe, the flux should be strongly suppressed (compared to its extrapolation from lower energies) as the effective volume within which sources can contribute shrinks.  The flux measurements by TA and Auger are shown in figure \ref{fig:spectra}, and compared to some astrophysical scenarios. 
Although the measurements of HiRes, Auger and TA are compatible within their systematic and statistical uncertainties \cite{SpectrumWG2012}, they are open to alternative interpretations. 

The TA Collaboration \cite{TAspectrum,Kido2013} finds their measurements at the highest end of the spectrum to be compatible with the GZK expectation, for a uniform distribution of sources that accelerate protons to energies well above the GZK threshold, as was also concluded by HiRes \cite{HiRes2008}. In models with a single extragalactic proton component, the ankle around $3\times 10^{18}$ eV can be interpreted as a propagation feature, a ``dip"  due to pair-production against the CMB \cite{Hill1985,Berezinsky1988,Berezinsky2005,Berezinsky2006}, and would not signal the transition between galactic and extragalactic dominance. TA measurements of the depth of shower maximum are compatible with a pure proton composition \cite{TAcomposition}. 

The fit of the spectrum measured by Auger with a model based on a single extragalactic proton component requires instead a cut-off in the injection energy around $10^{20}$ eV, and is not satisfactory if maximum energies are well above the GZK threshold \cite{Letessier2013}. Reasonably good fits to the end of the spectrum  with iron primaries are also possible, which is not surprising since they have comparable ``GZK horizons". The ``dip" however is a feature expected for protons only. Some differences in the suppression features for proton or iron spectra are due to a slightly different energy dependence of the horizon, and to the pile-up being significant for protons only. Good fits are also possible with a mixed composition, as we discuss further in the next section. Auger measurements of the depth of shower maximum with the fluorescence detector \cite{Augercomposition1,Augercomposition2}, consistent with independent estimators based on data from its ground array, indicate a trend from a light towards a heavier composition as the energy increases above approximately $3\times 10^{18}$ eV, 
with a relatively narrow range of mass values at a given energy. Note that the composition is derived using simulations based on hadronic interaction models that extrapolate cross-sections, inelasticities and multiplicities to energies larger than those accessible at human-made accelerators. The uncertainties inherent to these extrapolations constitute a very rich field of cross-link between CR and particle physics, beyond the already achieved indirect measurement of the proton-air cross section above LHC energies \cite{Augercrosssection}. It is conceivable that there could be new particle physics leading to an unexpected change in the hadronic interactions \cite{Allen1,Allen2}.

\section{Magnetic horizons and maximum acceleration scenarios}

Magnetic fields between the Earth and the CR sources clearly impact the observed distribution of arrival directions, since they deflect the trajectories of charged particles. They may also have an important role in shaping the spectrum and composition that we observe. If the propagation is turbulent, the path lengths are much longer than in the quasi-rectilinear regime, up to the point that particles may not be able to reach us over a Hubble time at rigidities below a threshold that depends on the source distance and the field parameters. A ``magnetic horizon" \cite{Lemoine2005} has been suggested as a possible source of low-energy suppression of an extragalactic proton component of CRs. Indeed, turbulent fields in the nG range and with coherence lengths of about 1 Mpc may induce a cutoff at values of $E/Z$ around and below a few times $10^{18}$ eV.  It is curious to note that in the extreme (and probably unrealistic) case that all sources are surrounded by turbulent fields as strong as 100 nG extending over several Mpc, distant sources would not contribute to the flux at Earth at all energies \cite{Deligny2004}, since the low-energy suppression would start at energies above the GZK threshold. 
\begin{figure}[H]
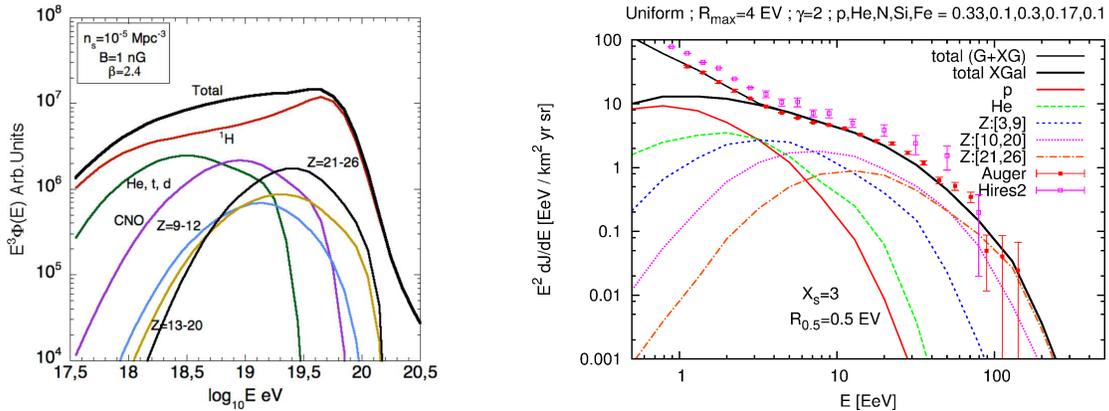

\centerline{\epsfig{file=allard.eps,width=2.25in}\hspace{.5in}\epsfig{file=diffusion.eps,width=3.in}}
\vspace*{8pt}
\caption{Flux rescaled by $E^3$ (left) or $E^2$ (right) predicted by scenarios with mixed composition injected at the sources with relatively high (left, taken from \cite{Allard2011}) and low (right, taken from \cite{Mollerach2013}) maximum rigidities, in the presence of turbulent magnetic fields.  In both cases the lines show the independent contributions to the observed spectrum of different mass groups. In the left panel the composition at the sources is assumed to be similar to the galactic composition, the spectral index is 2.4 and the maximum injection energy is $Z\times 10^{20.5}$ eV. The magnetic field has strength $B=1$ nG and coherence length $l_c=0.2$ Mpc, and the typical distance between sources is $d_s\approx 35$ Mpc. In the right panel the relative fractions are indicated and the maximum injection energy is $Z\times 10^{18.6}$ eV.  The sources spectral index is 2.0, and other parameters are such that $X_s\approx\frac{d_s}{\rm 65 Mpc}\sqrt{\frac{\rm Mpc}{l_c}}=3$ and $R_{0.5}\approx0.4\ {\rm EV} \frac{d_s}{\rm 100 Mpc}\frac{B}{\rm nG}\sqrt{\frac{\rm Mpc}{l_c}}=0.5$ EV.  Each mass group contributes to the flux in a relatively narrow energy range between the low energy magnetic cutoff and the maximum energies.  A ``galactic" component was added to match the total flux down to lower energies. 
\label{fig:mixedcomp}}
\end{figure}
One may think that the increased path length could also shape the spectral steepening at the highest energies, since it implies increased energy attenuation in the flux from a given source at a fixed rectilinear distance. However a propagation theorem \cite{Aloisio2004} guarantees that if the distribution of sources is continuous, then the total CR flux is the same as it would be in the absence of magnetic fields. The missing flux from relatively distant sources is exactly compensated by the increased containment of the flux diffusing from nearby sources. The same conclusion stands for a more realistic distribution of discrete sources, as long as the distance to the nearest ones is smaller than the diffusion length and the energy loss length. If the observer lies within the diffusion sphere of the nearby sources then the spectrum is unchanged compared to the case of an overall continuous distribution. The spectrum is affected only at rigidities such that the nearest sources are suppressed. 

These features are relevant in models that assume a mixed composition at the sources with rigidity-dependent acceleration limits, that may fit the highest energy end of the CR spectrum and also follow the trend in the mass composition suggested by Auger measurements. In scenarios with sufficient magnetic turbulence, a transition to a heavier composition above  $3\times 10^{18}$ eV with a relatively narrow range of mass values at a given energy would naturally result if protons and nuclei can no longer be accelerated above energies of the order of a few times  $Z \times 10^{18}$ eV \cite{Globus2008,Aloisio2011,Allard2011,Mollerach2013}. In these scenarios, each component contributes in the energy range between its low-energy magnetic cut-off and the maximum acceleration limit. If instead the maximum energy per unit charge at the sources reaches large values, of order  $10^{20.5}$ eV, an initially mixed composition at the source becomes gradually lighter during propagation, as a result of the photodisintegration of heavy nuclei \cite{Allard2005,Allard2011}. In models with composition dominated by intermediate mass nuclei at injection, it is  possible that all the light extragalactic component is due to the photodisintegration processes \cite{Hooper2010}. 

Examples of flux predictions in mixed-composition scenarios are illustrated in figure \ref{fig:mixedcomp}, for the case of maximum acceleration beyond the GZK threshold (left panel) and for lower maximum rigidities (right panel). One important difference between the scenario of a single extragalactic proton component and those with mixed composition is that in the latter, unless the proton fraction is high, there is no significant pair-production dip \cite{Allard2005,Aloisio2011}, and the explanation of the ankle requires a separate component, possibly of galactic origin.

\section{GZK smoking guns}

The existence of a light component in the CR flux at the highest energies is crucial if their arrival directions are to tell us about their place of origin. While protons with energies above the GZK threshold are expected to deviate by no more than a few degrees from a straight propagation in most parts of the sky, heavy nuclei are not likely to maintain a correlation of their arrival directions with their birthplaces, due to intervening magnetic fields. 
A ``smoking gun" indicative of energy-loss propagation effects acting upon a light component of CRs  would be a significant correlation between their arrival directions at energies above the GZK threshold and the directions of nearby extragalactic objects. 

\begin{figure}[H]
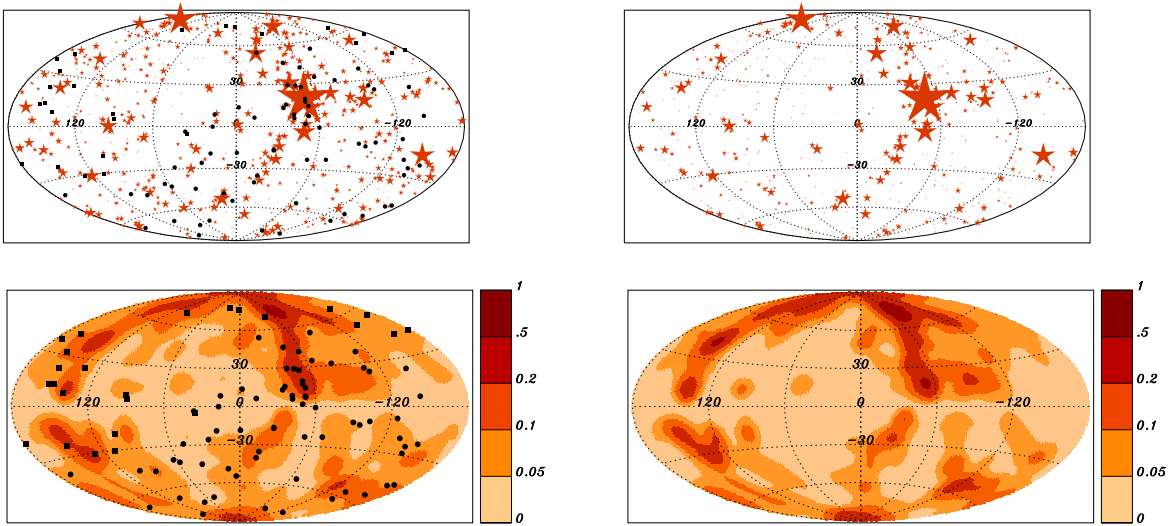

\centerline{\hspace{-.2in}\epsfig{file=bat70.gzk60.ps,width=2.5in}\hspace{.75in}\epsfig{file=bat70.gzk80.ps,width=2.5in}}
\vspace{-.75in}
\centerline{\epsfig{file=2mrsgzk60.ps,width=3.25in}\epsfig{file=2mrsgzk80.ps,width=3.25in}}
\vspace{-18pt}
\caption{Top panels: AGNs in the 70-month Swift-BAT catalog, displayed as stars with an area proportional to their X-ray flux, taking into account the attenuation factor expected from the GZK effect for energy thresholds of $6\times 10^{19}$ eV (left) and  $8\times 10^{19}$ eV (right). 
Bottom panels: $5^\circ$-smoothed density maps derived from 2MRS galaxies weighted by their infrared flux and by the same GZK attenuation factors as in the top panels. Maps are in galactic coordinates. As the energy threshold increases, fewer objects dominate the scene. Arrival directions of 25 events with energy above $5.7\times 10^{19}$ eV  measured by TA (filled squares) \cite{TA25events} and 69 events with energy above $5.5\times 10^{19}$ eV measured by Auger (filled circles) \cite{update2010}, are shown for illustration in the left panels. Note that these maps are not weighted by the relative exposure of a particular Observatory.
\label{fig:bat70}}
\end{figure}

The distribution of extragalactic matter within the GZK horizon is inhomogeneous. Arrival directions of CRs with nearby extragalactic origin should be more correlated with the matter distribution than an isotropic particle flux, if the trajectories are not significantly deflected by galactic or intergalactic magnetic fields. As an illustration, in figure \ref{fig:bat70} (top panels) we show all the active galactic nuclei (AGNs) in the 70-month Swift Burst Alert Telescope (BAT) catalog \cite{BAT70}, displayed as stars with an area proportional to their X-ray flux, taking into account the attenuation factor expected from the GZK effect upon protons for energy thresholds of $6\times 10^{19}$ eV (left) and  $8\times 10^{19}$ eV (right). Similarly, in the bottom panels we show density maps, smoothed over an angular scale of $5^\circ$, derived from the galaxies in the 2MASS Redshift Survey (2MRS) \cite{2MRS} weighted by their infrared flux and by the GZK attenuation factor. In both cases it is evident that the higher the energy threshold, the fewer objects dominate the scene. If the arrival directions of CRs followed a pattern similar to what is illustrated here, it would constitute firm additional circumstantial evidence for the action of the GZK effect.

Using data collected through 31 August 2007 the Pierre Auger Collaboration reported evidence for anisotropy in the distribution of the arrival directions of CRs with energy above $\approx  5.5\times 10^{19}$ eV \cite{Science2007,correlations2008}. The arrival directions were found to correlate within angular separations smaller than  $3.1^\circ$ with the positions of AGNs within 75 Mpc from the catalog by V\'eron-Cetty and V\'eron (VCV) \cite{VCVcatalog}. A test with independent data established a confidence level of 99\% for the rejection of the isotropic hypothesis.  The region of the sky close to the location of the radiogalaxy Cen A gave the largest observed excess with respect to isotropic expectations \cite{update2010}. Updates of these analyses \cite{update2011} did not increase the evidence for anisotropy. The pattern of arrival directions measured by Auger was also compared with that of 2MRS galaxies, and with AGNs detected by Swift-BAT \cite{update2010}. There is no statistically significant evidence that would favor a specific astrophysical scenario. 

Arrival directions of the events with highest energies measured by TA have also been compared with the VCV and other catalogs of nearby extragalactic objects \cite{TAcorrelations}. The smallest chance probability,  of order 1\% accounting for the scan in energy, angular separation and redshift, 
 was found with the Swift-BAT (58-month) AGN catalog. Inspection of the event distribution suggests that the deviations from isotropy hinted at by this and other tests may result mostly from a concentration of events in a $20^\circ$ region not far from the supergalactic plane, but the statistical significance of current observations is not sufficient to exclude a fluctuation.
Since the energy lost by CRs in their interaction with the CMB goes into gamma rays and neutrinos, another ``GZK smoking gun" would be the measurement of a diffuse flux of photons and neutrinos with ultra-high energies. This would add a significant piece to the puzzle of the CR origin. Bounds on the photon flux \cite{augerphotons1,augerphotons2} are approaching the GZK expectations \cite{Gelmini2008}. The cosmogenic neutrino flux is very dependent on the composition of the CR beam, and on the cosmological evolution of the sources. Current bounds \cite{icecubenus,augernus} are close to the flux expected in some models. The recent detection by IceCube \cite{icecubePeVnus} of neutrinos with energy above 60 TeV compatible with a cosmic origin is an interesting step forward. 
 
\section{Summary and conclusions}

The existence of a suppression of the flux of CRs at the highest energies has been confirmed, and it occurs within the energy range that was anticipated if the GZK effect were the cause. The spectral features, as measured with the current exposure, resolution, and systematic uncertainties of several experiments, are however insufficient to confidently establish whether the suppression is due to propagation effects acting on protons only, or else acting on a heavier and eventually mixed composition injected by extragalactic sources. It could also be the case that the suppression is not predominantly caused by propagation effects, if the sources (either extragalactic or galactic) happen to reach their maximum acceleration power at energies comparable to or just below the GZK threshold. This coincidence may seem unlikely, but in fact energies around $Z\times 10^{20}$ eV are at the imaginable limits of astrophysical accelerators. 

Current results of the Telescope Array are compatible with the GZK expectation for a uniform distribution of extragalactic sources that accelerate protons to energies well above the GZK threshold. In this scenario the ankle around $3\times 10^{18}$ eV can be interpreted as a propagation feature, instead of a signature of the transition between galactic and extragalactic dominance. 

The scenario based on extragalactic sources of protons with standard spectral index and accelerated to energies well beyond the GZK threshold does not provide a satisfactory fit to the measurements made with the  Pierre Auger Observatory. Maximum injection energies around $10^{20}$ eV are necessary for a reasonably good fit to this model. Additionally, Auger measurements of depth of shower maximum, interpreted according to current hadronic interaction models, suggest a heavier composition as the energy increases above approximately $3\times 10^{18}$ eV, with a relatively narrow range of mass values at a given energy. Flux measurements are compatible with scenarios such that the sources accelerate a mixed composition up to a maximum rigidity either above or below the GZK threshold. The combination of acceleration up to a maximum rigidity and a low rigidity cut-off due to diffusion across intervening magnetic fields could make each mass component contribute in a relatively narrow energy range. It is also possible that the sources inject a single mass component, and that the lighter elements arise as a result of photodisintegration during propagation.

At present there is no ``GZK smoking gun" from searches for anisotropy in the distribution of arrival directions at the highest energies. Either a significantly increased exposure with good energy resolution,
 or implementation of techniques that may allow a discrimination of CR composition on an event by event basis, may be needed for this type of exploration to bear fruit. 

Composition and anisotropy information, as well as evidence for cosmogenic neutrinos and photons produced during propagation, may be necessary in addition to spectral measurements to reach unambiguous conclusions about the origin of the suppression and to advance our understanding of the mechanisms behind the acceleration of CRs to energies around $10^{20}$ eV and beyond.



\section*{Acknowledgements}
I am grateful to the members of the Pierre Auger Collaboration for the chance to take part in this exciting project. I am indebted to Silvia Mollerach and Esteban Roulet for all the insight I gained from them during our longstanding collaboration on issues of CR propagation. I thank Stephane Coutu, Carola Dobrigkeit, Glennys Farrar and Paul Sommers for useful comments on the manuscript. This work was funded by PIP 01830 (CONICET) and PICT 1531 (ANPCyT).


\begin{thebibliography}{00}

\bibitem{Greisen1966} K. Greisen, Phys. Rev. Lett. {\bf 16} (1966) 748.

\bibitem{Zatsepin1966} G. T. Zatsepin and V. A. Kuz'min, JETP Lett. {\bf 4} (1966) 78.

\bibitem{Penzias65}A. A. Penzias and R. W. Wilson, Astrophys. J. {\bf 142} (1965) 419.

\bibitem{Linsley1963}J. Linsley, Phys. Rev. Lett. {\bf 10} (1963) 146.

\bibitem{HiRes2008}HiRes Collaboration, Phys. Rev. Lett. {\bf 100} (2008) 101101.

\bibitem{Auger2008}Pierre Auger Collaboration, Phys. Rev. Lett. {\bf 101} (2008) 061101.

\bibitem{Takeda2003}M. Takeda et al., Astropart. Phys. {\bf 19} (2003) 447.

\bibitem{Letessier2013}A. Letessier-Selvon for the Pierre Auger Collaboration, 33rd ICRC (2013)  arXiv:1310.4620. 

\bibitem{TAspectrum}Telescope Array Collaboration, Astrophys.J. {\bf 768} (2013) L1.

\bibitem{Kido2013}E. Kido for the Telescope Array Collaboration, 33rd ICRC (2013) arXiv:1310.6093.

\bibitem{SpectrumWG2012}B. Dawson et al., EPJ Web of Conferences {\bf 53} (2013) 01005.

\bibitem{Stecker1968} F. Stecker, Phys. Rev. Lett. {\bf 21} (1968) 1016.

\bibitem{Hill1985} C. Hill and D. Schramm, Phys. Rev. D {\bf 31} (1985) 564.

\bibitem{Berezinsky1988} V. Berezinsky and S. Grigorieva, Astron. \& Astrophys. {\bf 199} (1988) 1.

\bibitem{Berezinsky1990} V. Berezinsky, S. Grigorieva and V. A. Dogiel, Astron. Astrophys. {\bf 232}  (1990) 582.

\bibitem{Aharonian1990} F. Aharonian, B. Kanevsky and V. Vardanian, Astrophys. Space Sci. {\bf 167} (1990) 93.

\bibitem{Yoshida1993} S.  Yoshida and M. Teshima, Prog. Theor. Phys. {\bf 89} (1993)  833. 

\bibitem{Rachen1993} J. P. Rachen and P. Biermann, Astron. \& Astrophys. {\bf 272} (1993) 161.

\bibitem{Aharonian1994} F. Aharonian and J. Cronin, Phys. Rev. D {\bf 50} (1994) 1892.

\bibitem{Waxman1995} E. Waxman, Astrophys. J. {\bf 452} (1995) L1. 

\bibitem{Stanev2000}T. Stanev et al., Phys. Rev. {\bf D62} (2000) 093005.

\bibitem{Berezinsky2005} V. Berezinsky, A. Z. Gazizov and S. I. Grigorieva, Phys. Lett. {\bf B612} (2005) 147.

\bibitem{Berezinsky2006}V. Berezinsky, A. Z. Gazizov, and S. I. Grigorieva, Phys. Rev. {\bf D 74}  (2006) 043005.

\bibitem{Kachelriess2009}M. Kachelriess, E. Parizot and D.V. Semikoz, JETP Lett. {\bf 88} (2009) 553. 

\bibitem{Puget1976} J. L. Puget, F. Stecker and J. H. Bredekamp, Astrophys. J. {\bf  295} (1976) 638.

\bibitem{Epele1998} L. N. Epele and E. Roulet, JHEP {\bf 9810} (1998) 009.

\bibitem{Stecker1999} F. Stecker and M. H. Salamon, Astrophys. J. {\bf 512} (1999) 521. 

\bibitem{Khan2005} E. Khan et al., Astropart. Phys. {\bf 23} (2005) 191.

\bibitem{Allard2005} D. Allard et al., Astron. \& Astrophys. {\bf 443} (2005) L29.

\bibitem{Allard2006} D. Allard et al., JCAP {\bf 0609} (2006) 005.

\bibitem{Harari2006}D. Harari, S. Mollerach and E. Roulet, JCAP {\bf 0611} (2006) 012.

\bibitem{Hooper2007}D. Hooper, S. Sarkar and A. Taylor,  Astropart. Phys. {\bf 27} (2007) 199. 

\bibitem{Allard2008} D. Allard et al., JCAP {\bf 0810} (2008) 033.

\bibitem{Hooper2008} D. Hooper, S. Sarkar and A. Taylor,  Phys. Rev. {\bf D 77} (2008) 103007. 

\bibitem{Allard2009}D. Allard and R.J. Protheroe, Astron. \& Astrophys. {\bf 502} (2009) 803.

\bibitem{Watson2013}A. Watson,  arXiv:1310.0325 (2013).

\bibitem{Kotera2011}K. Kotera, A. V. Olinto, Ann. Rev. Astron. Astrophys. {\bf 49} (2011) 119. 

\bibitem{Stanev2009}T. Stanev, New J. Phys. {\bf 11} (2009) 065013.

\bibitem{Kampert2013}K.-H. Kampert et al., Astropart. Phys. {\bf 42} (2013) 41.

\bibitem{Aloisio2012}R. Aloisio et al., JCAP {\bf 1210} (2012) 007.

\bibitem{Berezinsky1969}V.S. Berezinsky and G.T. Zatsepin, Phys. Lett. {\bf B 28} (1969) 423.

\bibitem{Stecker1979}F. W. Stecker, Astrophys. J. {\bf 238} (1979) 919.

\bibitem{Engel2001}R. Engel, D. Seckel, T. Stanev, Phys. Rev. {\bf D 64} (2001) 093010.

\bibitem{Hooper2005}D. Hooper, A. Taylor, S. Sarkar, Astropart. Phys. {\bf 23} (2005) 11. 

\bibitem{TAcomposition}Y. Tsunesada for the Telescope Array Collaboration, 33rd ICRC (2013).

\bibitem{Augercomposition1}Pierre Auger Collaboration, Phys. Rev. Lett. {\bf 104} (2010) 091101.

\bibitem{Augercomposition2}Pierre Auger Collaboration, JCAP {\bf 02} (2013) 026. 

\bibitem{Augercrosssection}Pierre Auger Collaboration, Phys. Rev. Lett. {\bf  109} (2012) 062002.

\bibitem{Allen1}G. R. Farrar and J. D. Allen, European Physical Journal Web of Conferences {\bf 53} (2013) 7007.
    
\bibitem{Allen2}J. Allen and G. Farrar, arXiv:1307.7131 (2013). 

\bibitem{Lemoine2005}M. Lemoine, Phys. Rev. {\bf D71} (2005) 083007.

\bibitem{Deligny2004}O. Deligny, A. Letessier-Selvon and E. Parizot, Astropart. Phys. {\bf 21} (2004) 609. 

\bibitem{Aloisio2004}R. Aloisio and V. Berezinsky, Astrophys. J 612 (2004) 900.

\bibitem{Globus2008} N. Globus, D. Allard and E. Parizot Astron. \& Astrophys. {\bf 479} (2008) 97.

\bibitem{Aloisio2011}R. Aloisio, V. Berezinsky and A. Gazizov,  Astropart. Phys. {\bf 34} (2011) 620.

\bibitem{Allard2011} D. Allard,  Astropart. Phys. {\bf 39-40} (2012) 33 (extended version available in arXiv:1111.3290).

\bibitem{Mollerach2013}S. Mollerach and E. Roulet, JCAP {\bf 1310} (2013) 013.

\bibitem{Hooper2010} D. Hooper and A. Taylor, Astropart. Phys. {\bf 33} (2010) 151.

\bibitem{TA25events}Telescope Array Collaboration, Astrophys. J. {\bf 757} (2012) 26. 

\bibitem{update2010}Pierre Auger Collaboration, Astropart. Phys. {\bf 34} (2010) 314.

\bibitem{BAT70}W.H. Baumgartner et al., Astrophys. J. Suppl. {\bf 207} (2013) 19.

\bibitem{2MRS}J. P. Huchra et al., Astrophys. J. Suppl. {\bf 199} (2012) 26.

\bibitem{Science2007}Pierre Auger Collaboration, Science {\bf 318} (2007) 938.

\bibitem{correlations2008}Pierre Auger Collaboration, Astropart. Phys. {\bf 29} (2008) 188.

\bibitem{VCVcatalog}M.-P. V\'eron-Cetty and P. V\'eron, Astron. \& Astrophys. {\bf 455} (2006) 773.

\bibitem{update2011}K.-H. Kampert for the Pierre Auger Collaboration, 32nd ICRC (2011)  arXiv:1207.4823.

\bibitem{TAcorrelations}Telescope Array Collaboration,  Astrophys. J. {\bf 777} (2013) 88.

\bibitem{augerphotons1}Pierre Auger Collaboration, Astropart. Phys. {\bf 29} (2008) 243.

\bibitem{augerphotons2}Pierre Auger Collaboration, Astropart. Phys. {\bf 31} (2009) 399.

\bibitem{Gelmini2008}G. Gelmini, O. E. Kalashev and D. V. Semikoz, JETP {\bf 106} (2008) 1061.

\bibitem{icecubenus}IceCube Collaboration, arXiv:1311.7048 (2013).

\bibitem{augernus}Pierre Auger Collaboration, Astrophys. J. {\bf 755} (2012) L4. 

\bibitem{icecubePeVnus}IceCube Collaboration, Science {\bf  342} (2013) 6161.

\end{thebibliography}
\end{document}